\documentclass[conference]{IEEEtran}
\IEEEoverridecommandlockouts
\usepackage{cite}
\usepackage{makecell}
\usepackage{soul}
\usepackage{amsmath,amssymb,amsfonts}
\usepackage{algorithm,algorithmic}
\usepackage{graphicx}
\usepackage{textcomp}
\usepackage{xcolor}

\def\BibTeX{{\rm B\kern-.05em{\sc i\kern-.025em b}\kern-.08em
    T\kern-.1667em\lower.7ex\hbox{E}\kern-.125emX}}

\begin{document}

\title{Clustering Interval Load with Weather to Create Scenarios of Behind-the-Meter Solar Penetration
\\

\thanks{This work was performed at the Pacific Northwest National Laboratory, operated by the Battelle Memorial Institute under the auspices of the U.S. Department of Energy (Contract DE-AC05-76RL01830), supported by the Solar Energy Technologies Office (SETO) project no. 37774. }
}


\author{\IEEEauthorblockN{Allison M.\ Campbell,\IEEEauthorrefmark{1} Soumya Kundu,\IEEEauthorrefmark{1} Andrew Reiman,\IEEEauthorrefmark{1} Orestis Vasios,\IEEEauthorrefmark{1} Ian Beil,\IEEEauthorrefmark{2} and Andy Eiden\IEEEauthorrefmark{2}}
\IEEEauthorblockA{\IEEEauthorrefmark{1} Pacific Northwest National Laboratory, USA\\
Email:\,\{allison.m.campbell,\,soumya.kundu,\,andrew.reiman,\,orestis.vasios\}@pnnl.gov}
\IEEEauthorblockA{\IEEEauthorrefmark{2} Portland General Electric, USA\\
Email:\,\{ian.beil,\,andy.eiden\}@pgn.com}}

\maketitle

\begin{abstract}
Forecasting load at the feeder level has become increasingly challenging with the penetration of behind-the-meter solar, as this self-generation (also called total generation) is only visible to the utility as aggregated net-load. This work proposes a methodology for creation of scenarios of solar penetration at the feeder level for use by forecasters to test the robustness of their algorithm to progressively higher penetrations of solar. The algorithm draws on publicly available observations of weather \emph{condition} (e.g., rainy/cloudy/fair) for use as proxies to sky clearness. These observations are used to mask and weight the interval deviations of similar native usage profiles from which average interval usage is calculated and subsequently added to interval net generation to reconstruct interval total generation. This approach improves the estimate of annual energy generation by 23\%; where the net generation signal currently only reflects 52\% of total annual generation, now 75\% is captured via the proposed algorithm. This proposed methodology is data driven and extensible to service territories which lack information on irradiance measurements and geo-coordinates.

\end{abstract}


%


\section{Introduction}
Behind-the-Meter (BTM) solar installations have grown precipitously in recent years. In the United States, rooftop solar installations grew by 51\% between 2020 and 2021 and are projected to continue with modest growth to 41 GW annually by 2026 \cite{solaroutlook2022}. These installations allow utility customers to self-provide power, which fundamentally alters the load profile measured by Advanced Metering Infrastructure (AMI) equipment. This new load profile measured at each customer meter now represents the net-load -- the customer's native interval usage (negative) plus interval self generation (positive), which caps at zero net-load when generation exceeds demand.  In response to the changing content of the net-load signal, grid operators and planners must incorporate additional information into their load forecasting algorithms to account for multiple parameters: (1) Does the customer have rooftop solar? (2) How large is the system? (3) What are the system configuration parameters? (e.g., panel tilt and orientation, inverter size) and (4) What is the available solar irradiance at each location?

Utilities typically only know (1) as customers with rooftop solar request net metering tariffs. Identification of items (2), (3), and (4) relies on spotty registration data and low spatial coverage of meteorological measurements. In practice, efforts to accurately forecast net-load are focused on filling in the gaps for these intervening parameters to facilitate high-resolution modeling of disaggregated generation \cite{wang2018}. For utilities that either do not have the capacity to incorporate sophisticated estimation algorithms for these parameters to forecast BTM solar generation and native load separately, or have developed net-load forecasting algorithms which side step these values, the ultimate test of the algorithm is its robustness to increasing penetration of BTM solar.

We define the penetration of BTM solar as the total annual generation of solar for all customers in a region managed by a utility divided by the total annual native demand. This definition deviates from previous studies \cite{erdener2022} to emphasize the role of the aggregated interval \emph{generation} as opposed to the \emph{capacity} for generation. Particularly in regions with highly variable solar irradiance from moving clouds and other weather, framing the penetration of solar in terms of capacity will necessitate high spatial resolution irradiance measurements. The skill of a net-load forecast algorithm will be more accurately measured as a function of the total BTM generation.


The rest of the paper is organized as follows. Section \ref{sec:back}
states the problem. Section \ref{sec:method} proposes the framework and
details of the methodology. In Section \ref{sec:results}, results are discussed.
Finally, conclusions are presented in Section \ref{sec:concl}.

\section{Background}\label{sec:back}

The literature on BTM solar generation is largely focused on disaggregation of interval solar generation from net-load measured at a single meter. A review of disaggregation methods applied to BTM solar can be found in \cite{erdener2022}, wherein the majority of approaches rely on a physical solar model (PSM) to calculate solar power independent of load.  The purpose of estimating installed capacity is to create an independent series of BTM production from the PSM and add it to a separate model of native demand \cite{sun2020}. The estimation of capacity can be accomplished though sampling an empirical distribution of rooftop solar configuration parameters \cite{saint-drenan2018}, however this relies on knowledge of at least a subset of installed capacities registered at the household level, which are partially known at best \cite{jin2021,sun2020}. Authors of \cite{chen2021} note that the lack of system configuration specifics is the largest barrier to estimation of interval generation.  These approaches assume that native load and BTM solar generation must be estimated independently.

Development of statistically informed BTM interval generation has attempted to fill two measurement gaps: physical system characteristics and available global horizontal irradiance (GHI). Statistical estimation of interval generation has been proposed to bypass PSM. The work in \cite{quilumba2015} identified similar customer usage profiles using k-means clustering to forecast customers with and without rooftop solar separately. In \cite{shaker2016a}, the authors implemented a hybrid k-means clustering and principal component analysis to identify BTM solar customers, trained with a subset of known solar customer profiles and known distance between locations. This approach was extended in \cite{shaker2016b} to use weather to define regions of similar generation. The primary limitation of this approach is the need for geo-coordinates, which are often held private by the utility to preserve customer privacy. The authors in \cite{alam2020} incorporated high spatial resolution measurements of GHI to cluster distinct weather patterns on BTM PV generation, acknowledging the impact of cloud formation variability on forecast uncertainty.


The goal of these studies was to reconstruct interval generation through reliance on high resolution data. The PSM approach requires PV system characteristics while statistical models require geo-coordinates and GHI.  This work proposes to improve the reconstruction of BTM solar generation by relying only on publicly available weather conditions.  The purpose of this methodology is to improve customer-level estimation of annual generation for use in developing synthetic scenarios of high BTM solar penetration against which net-load forecast algorithms can be tested. 

The contributions of this work are as follows. (1) A methodology is proposed to create scenarios of BTM solar penetration for a utility feeder based on synthesized BTM solar generation and native load. (2) The synthesized BTM solar generation is calculated for each solar customer through comparison with similar usage profiles from non-solar customers. (3) A native load similarity weight is applied as a function of the prevailing cloudiness throughout each day; usage profiles between customers with and without solar are deemed ``similar'' on overcast days. (4) The recaptured BTM solar generation is used to improve the estimate of BTM solar penetration and to create customer aggregation scenarios against which net-load forecast algorithms can be tested.

\section{Methodology}\label{sec:method}

Load and generation time series data for this work come from AMI meters within the Portland General Electric (PGE) territory.  PGE has two types of programs for customers with rooftop solar: (1) Solar net-meter, where one meter records net-load (capped at zero) and another meter records net-generation in an interval, and (2) Solar Purchase Option (SPO), which provides signals for net-load, net-generation, and total interval generation. The PGE dataset is comprised of three types of customers: (1) customers without solar, (2) solar customers with a net-metering agreement, and (3) solar customers with SPO and net-metering agreement.

\begin{figure}[thpb]
\begin{center}
\includegraphics[width = 1\linewidth]
{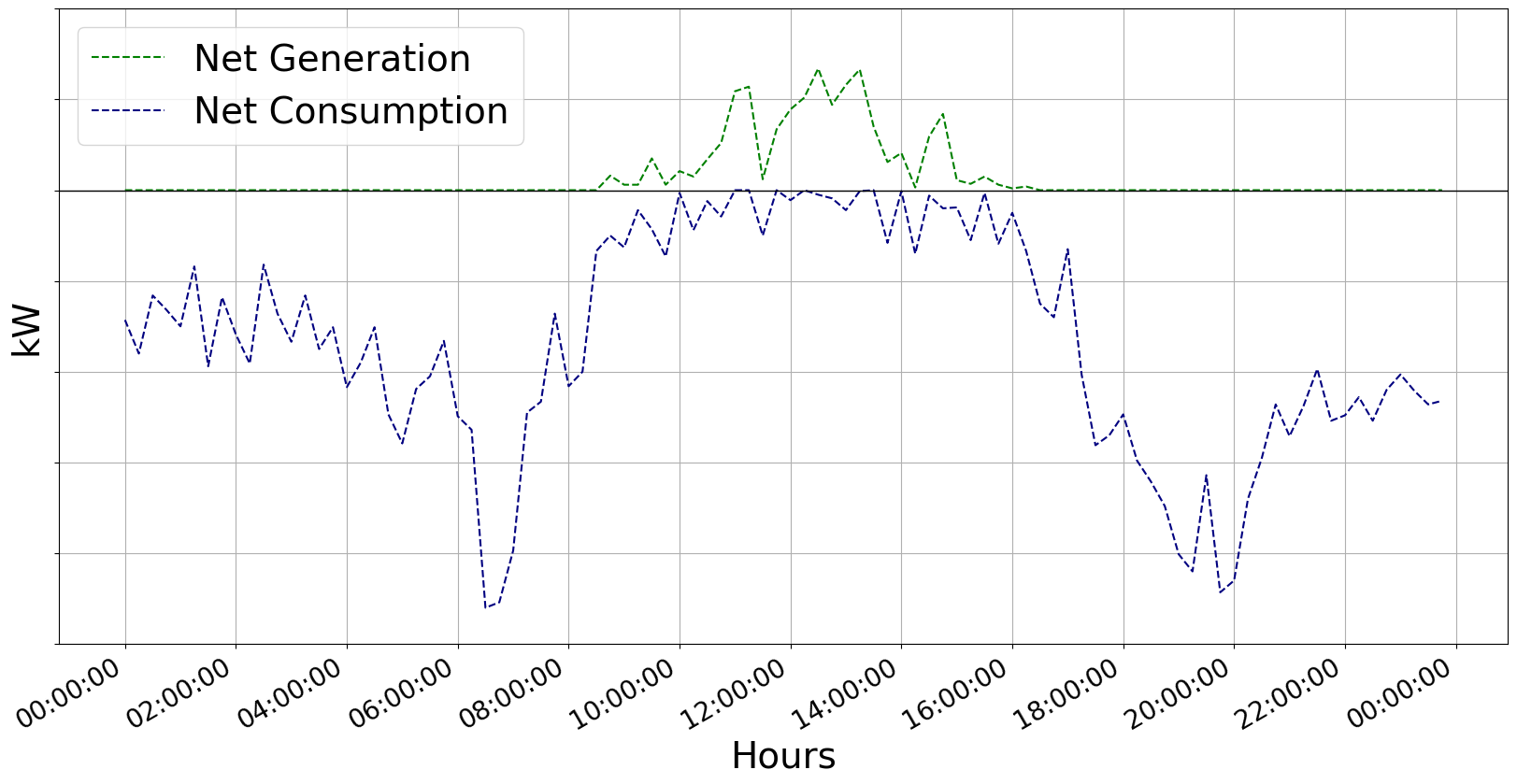}
\includegraphics[width=1\linewidth]{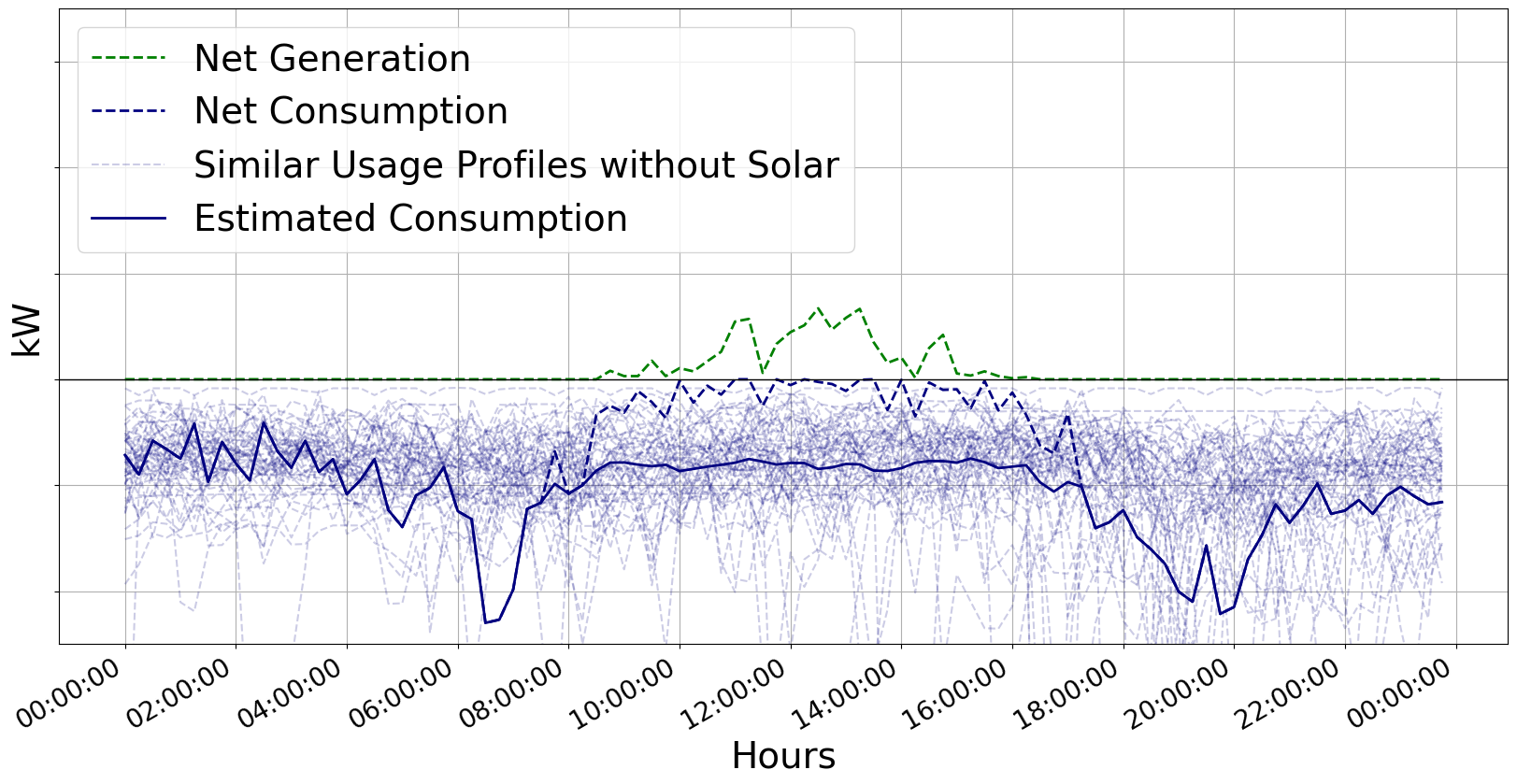}
\includegraphics[width=1\linewidth]{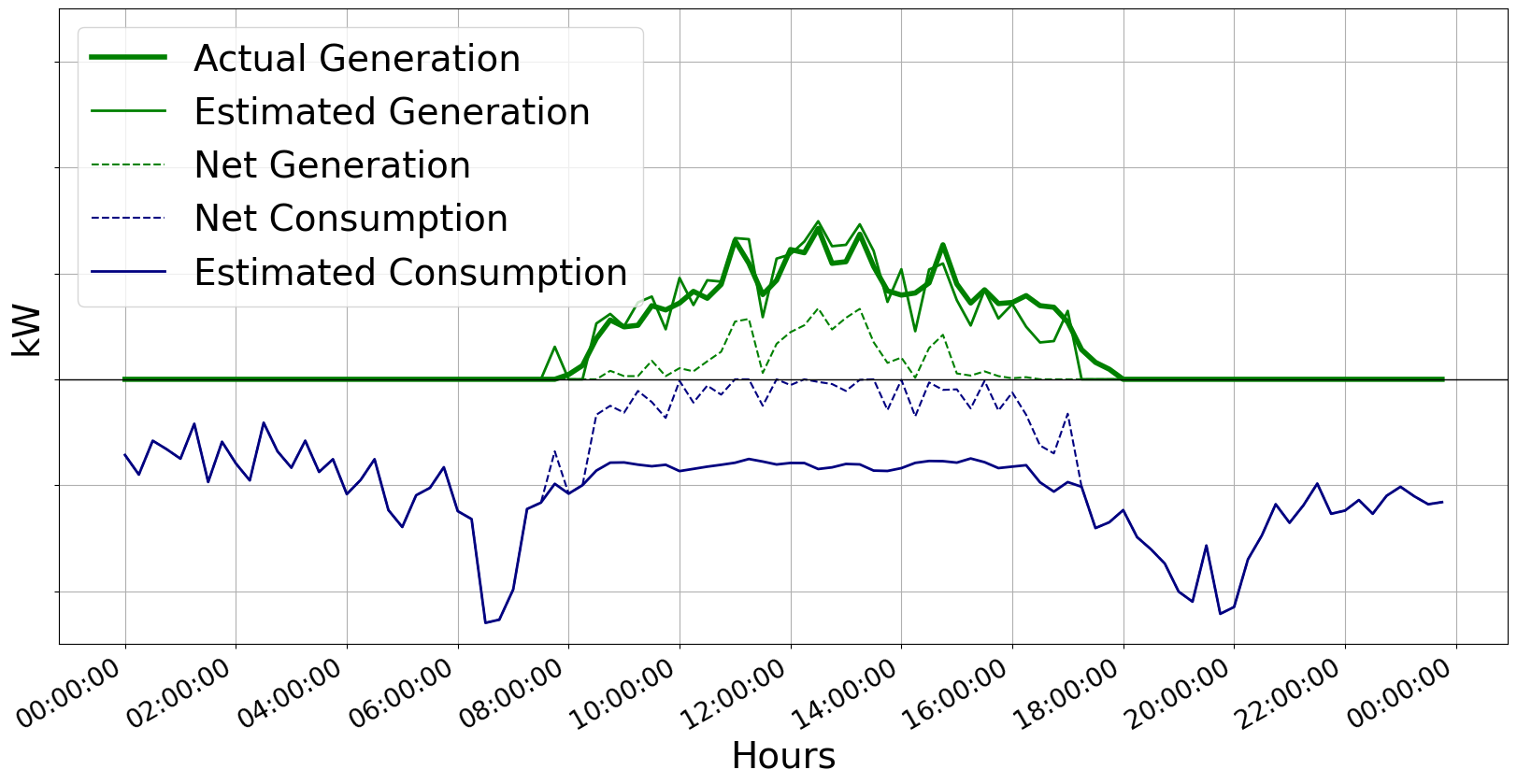}

    \caption{(Top) Net generation (green) and the net consumption (blue) of one customer. (Middle) Interval consumption of 75 similar non-solar customers. Light blue lines are individual non-solar customer usage. (Bottom) Extra generation identified by this approach effectively recaptures interval generation. Exact kW values on y-axis are removed to retain anonymity. }
    \vspace{-0.2in}
        \label{fig:one-spid}
\end{center}
\end{figure}

The dataset provided by PGE contained net consumption time series for 24,438 customers inclusive of the three customer types. Across sub-types, 63\% are without solar, 36\% are net-meter only, and 1.6\% are net-meter and SPO. The break-out of customer counts and usage/generation is shown in Table \ref{tab:PGE-cust}. Fig.\ \ref{fig:one-spid} (Top) illustrates the anonymized net generation (excess generation exported to the grid) and net consumption (amount billed to the customer) for one customer over 24 hours. 


\begin{table}[thpb]
\caption{PGE residential customer counts and annual energy. }\label{tab:PGE-cust}
\centering
\begin{tabular}{lll}
Utility Dataset            & Count  & Annual Sum \\ \hline \hline
No Solar   & 15,289 & 140 GWh    \\\hline 
Net-meter Net   Generation  & 7,862  & 28 GWh     \\\hline 
Net-meter Net   Consumption & 8,766  & 73 GWh     \\\hline 
SPO Net   Generation       & 369    & 1.4 GWh    \\\hline 
SPO Net   Consumption      & 383    & 5 GWh      \\\hline 
SPO Total Generation       & 383    & 2.5 GWh   
\end{tabular}
    \vspace{-0.2in}
\end{table}


Net-load forecasting accuracy is  a function of the impact of interval generation on the native load. Solar generation measurements are only available for 383 customers and many of these customers have sized their rooftop PV systems to meet their annual demand; this translates to a depressed percentage of household-level solar penetration by annual energy self-produced. Based on net generation and net consumption of all  383 SPO customers, the solar penetration is $\sim 28\%$. In order to reach higher penetration levels upwards of 50\% to test the robustness of net-load forecast algorithms, this methodology must reconstruct the additional BTM generation of net-meter solar customers. These customer profiles are then recombined to produce synthetic feeders with high BTM solar generation.




\subsection{Weather Information} 
Solar  generation depends on reliable solar irradiance; days with complete blockage due to clouds will proportionally yield reduced power,  those  devoid of clouds will yield full power, and those with varied conditions such as scattered clouds or smoke are known to yield rapid power fluctuations.  While cloud coverage on varied-condition days changes across the extent of a balancing authority (PGE is $\sim$60 miles across), an average-sized cloud can completely eclipse a distribution feeder; all BTM  generation in the feeder will experience similar reduced output from the passing cloud. 

We reconstruct the average BTM generation on days with clear weather by looking to the native usage of similar non-solar customers on overcast days. Averaging the profiles of customers with similar usage reduces downstream variance in marginal PV generation estimation which could arise from variations in PV installation status and equipment degradation. The goal is to estimate the average marginal native load for the target customer under clearsky conditions.

Usage profiles from non-solar customers are deemed similarity candidates to those of SPO customers on days when consumption is unlikely be impacted by BTM solar generation or when GHI is very low. Weather will be the dominant indicator of PV production; on rainy days PV output can be reduced by 40\% and on overcast days by 45\% \cite{gupta2019}.
 Interval cloud cover condition is used as a proxy for GHI with condition observations taken from the Portland International Airport \cite{wunderground}. 
Weather conditions and their similarity weighting (denoted by S) is shown in Table \ref{tab:weather_cond}. 
 Weights are assigned to a simplified weather condition, such as $S = 1$ (rainy), $S=0.1$ (cloudy), and $S=0.01$ (fair) to logarithmically preference the similarity of non-solar interval consumption during rainy intervals over others. The specific weight values may be tuned for improved accuracy.


\begin{table}[htpb]
    \caption{Weather conditions observed at the Portland Intl.\ Airport.}
    \label{tab:weather_cond}
    \centering
    \begin{tabular}{l|l|l}
         {Weather Condition}  & {Group} & {Similarity Weight (S)} \\ \hline \hline
Heavy   Rain, Rain   & Rainy  & 1    \\ \hline
\makecell[l]{Light   Rain, (Mostly/Partly) Cloudy,\\
(Shallow/Patches of) Fog, Mist, \\
Haze, Light Drizzle, Wintry Mix, \\ 
Smoke, Heavy Thunder-Storm} & {Cloudy} & {0.1} \\ \hline
Fair,   Light Snow, Windy, Thunder   & Fair  & 0.001   
    \end{tabular}

\end{table}

\subsection{Estimation of BTM Solar Generation}


We leverage non-solar customer profiles to reconstruct native SPO load on overcast days.  This approach is similar to \cite{liu2020}, which grouped customers based on usage profiles before estimation of BTM generation. There are no direct GHI measurements to reconstruct generation; in its place, recorded weather conditions (rainy/cloudy/fair) at the nearest weather station are used as proxy for sky clearness (Table \ref{tab:weather_cond}).  Testing of the algorithm is done with SPO customers supplemented with usage profiles from non-solar customers. The algorithm to complete this task for each SPO customer has three steps: 

\newcommand{\step}[2]{\noindent\textbf{Step #1:} \textit{#2}.}

\step{1}{Calculate Daily Averaged Similarity Weight ($\overline{S}_d$)} Considering only the daytime intervals (i.e., sunrise to sunset) containing a total of $m$ time-intervals $\lbrace t_1,t_2,\dots,t_m\rbrace$, the daily averaged similarity weight for the feeder region is obtained as:
\begin{align}
\label{eq:step1}
    \overline{S}_d &= \frac{ S_{t_1}+S_{t_2}+\dots+S_{t_m}}{m} &\text{(for each day $d$)}
\end{align}
where $S_{t}$ denotes the similarity weight at interval $t$\,.

\begin{figure}[thpb]
\begin{center}
\includegraphics[width = 1\linewidth]{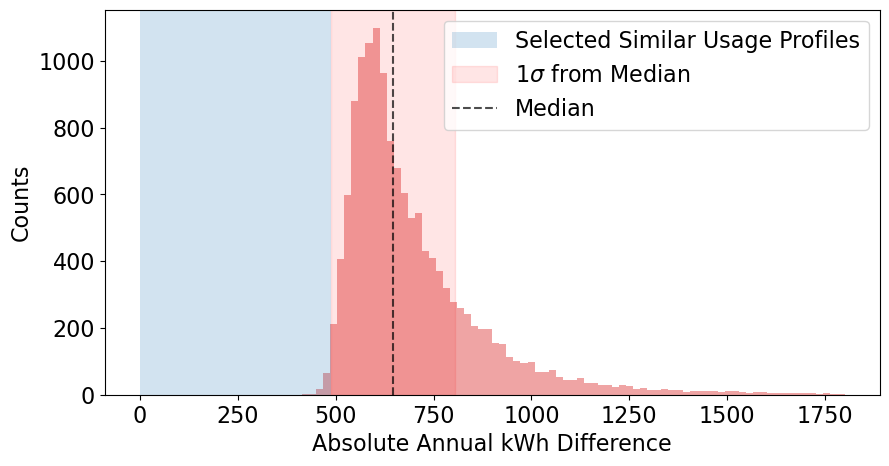}
\vspace{-0.2in}
\caption{Eligible similar customers are those whose absolute interval weighted difference lies below one standard deviation from the median.}
\label{fig:knn}
\end{center}
\end{figure}

\step{2}{Select Similar Non-Solar Load Profiles} For every solar customer $i$, we define the interval difference between their net usage profile ($u_{i,t}$) and the usage profile ($u_{j,t}$) of every non-solar customer $j$ as follows:
\begin{align}
\label{eq:step2_delta_t}
    \Delta_{i,j,t}&:=\left|u_{i,t}-u_{j,t}\right|\,, &\begin{array}{l}
         \forall i: \text{\{solar customers\}}  \\
          \forall j: \text{\{non-solar customers\}}\\
          \forall t: \text{\{time-intervals\}}
    \end{array}
\end{align}
Considering the $m$ day-time intervals $\lbrace t_1,t_2,\dots,t_m\rbrace$, the daily usage difference between the $i$-th solar customer and the $j$-th non-solar customer is obtained as:
\begin{align}
\label{eq:step2_delta_d}
    \Delta_{i,j,d} &= \Delta_{i,j,t_1} + \Delta_{i,j,t_2}+\dots+\Delta_{i,j,t_m} &\forall d:\text{\{days\}}
\end{align}
Next, the daily usage differences are weighted by the daily averaged similarity weights ($\overline{S}_d$) and summed over a full year to obtain the weighted annual usage difference between the $i$-th solar customer and the $j$-th non-solar customer as follows:
\begin{align}
\label{eq:step2_delta}
    \Delta_{i,j}=\sum_{d\in\text{\{days in year\}}}\left(\overline{S}_d\,\cdot\,\Delta_{i,j,d}\right)
\end{align}
Note that small $\Delta_{ij}$ suggests that the usage (net consumption) profile of the $j$-th non-solar customer is similar to the $i$-th solar customer. The empirical distribution of the $\Delta_{ij}$ values are used to identify the non-solar similar customers as:
\begin{align}
    \forall i:\quad \mathcal{N}_i&:=\left\lbrace j\,\left|\,\Delta_{ij}\leq \text{med}\left(\Delta_{ij}\right)\!-\!\sigma\left(\Delta_{ij}\right)\right.\right\rbrace\,,
\end{align}
where med$(\cdot)$ and $\sigma(\cdot)$ denote median and standard deviation, respectively. Only the customer profiles that are in the tail of the distribution -- at least one standard deviation below the median -- are considered as similar. Fig. \ref{fig:knn} shows that the non-solar customers with annual interval difference less than roughly 500 kWh are deemed most similar. For this SPO customer, this identifies roughly 75 non-solar customers whose annual consumptions on rainy or overcast days are the closest.

\step{3}{Correct Load and Generation on Fair Days} For each solar customer $i$, we have access to the net-metered values:
\begin{align*}
    u_{i,t}\leq 0:&\text{ net consumption}\,,\,~\,v_{i,t}\geq 0:\text{ net generation\,.}
\end{align*}
The goal is to estimate the total generation ($\hat{v}_{i,t}$), by adjusting the net-generation values with the help of the estimated total consumption ($\hat{u}_{i,t}$), calculated as follows:
\begin{align*}
    \hat{u}_{i,t}&\!=\!\frac{1}{\left|\mathcal{N}_i\right|}{\sum_{j\in\mathcal{N}_i}\!u_{j,t}}\,,
\end{align*}
where $\left|\mathcal{N}_i\right|$ denotes 
the number of similar non-solar customers. Finally, the total generation ($\hat{v}_{i,t}$) is estimated by adding a non-negative correction ($w_{i,t}\!\geq\! 0$) to its net generation ($v_{i,t}$)\,:
\begin{align}
\label{eq:step3}
    \hat{v}_{i,t} &= v_{i,t}+w_{i,t}\,,&\text{where }\,w_{i,t}\!:=\!\max(0,u_{i,t}\!-\!\hat{u}_{i,t})
\end{align}

The interval consumption of these similar non-solar customers is shown in Fig.\ \ref{fig:one-spid} (Middle) for a sunny day. The customer's usage was in excess of generation multiple times mid-day. The total consumption is replaced with the average of similar non-solar customers' usage, illustrated by the solid blue line in Fig.\ \ref{fig:one-spid} (Bottom).

\section{Results}\label{sec:results}

\begin{figure}[thpb]
\begin{center}
\includegraphics[width = 0.4\textwidth]{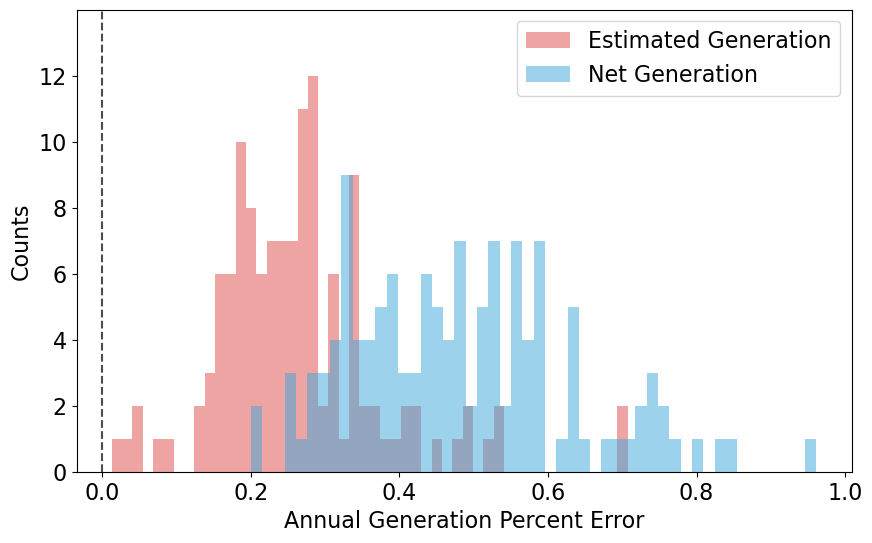}
\vspace{-0.1in}
\caption{Customer-level percent error in annual generation for estimated (red) and net provided at the meter (blue).}
\vspace{-0.2in}
\label{fig:valid_hist}
\end{center}
\end{figure}

Validation of the BTM solar generation reconstruction was done with SPO customers using their actual generation meter readings as comparison. Across these customers, the net generation series only accounts for $52\%$ of the total annual generation. Reconstitution of the interval generation using the approach in this work increases the representation of BTM generation to $75\%$ of total annual generation. The error in the annual generation can be seen in Fig.\ \ref{fig:valid_hist}, where the percent error for each customer is the annual actual generation minus net (or estimated) generation divided by actual. The error in the net generation is $48\%$, with a standard deviation of $15\%$. After application of this approach, the error in the estimated generation is $26\%$ with a standard deviation of $11\%$.

\begin{figure}[thpb]
    \centering
    \includegraphics[width=1\linewidth]{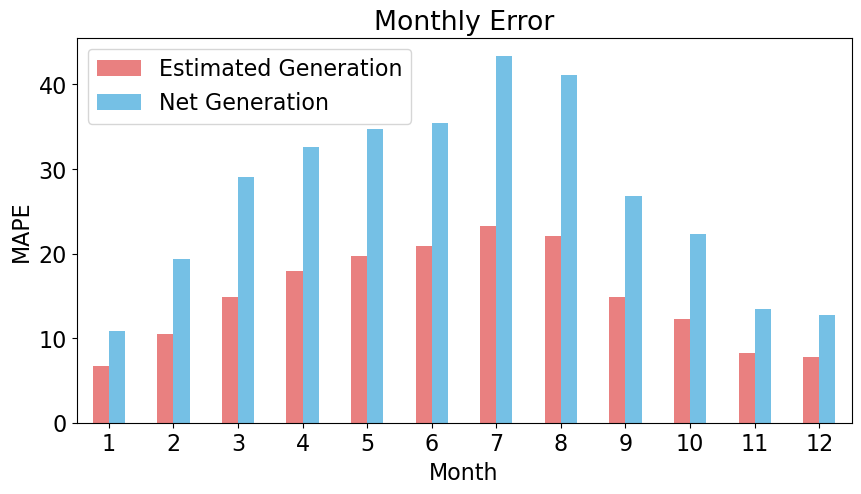}
    \vspace{-0.2in}
    \caption{Averaged across validation customers, the monthly MAPE is lower for the proposed approach to estimate BTM generation compared with the metered net generation.}
    \label{fig:monthly-mape}
\end{figure}

 The mean absolute percent error (MAPE) in monthly energy averaged across all SPO customers is shown in Fig.\ \ref{fig:monthly-mape}. Where Fig.\ \ref{fig:valid_hist} showed the annual error, this dives into the seasonal dependence. We see that this reconstruction approach improves the estimate of BTM solar generation in the summer months most. This approach selected non-solar customer usage profiles based on the integrated usage difference on overcast or rainy days; this allows us to leverage the same usage profiles to reconstruct the native demand on sunny days.

\begin{figure}[thpb]
    \centering
    \includegraphics[width=1\linewidth]{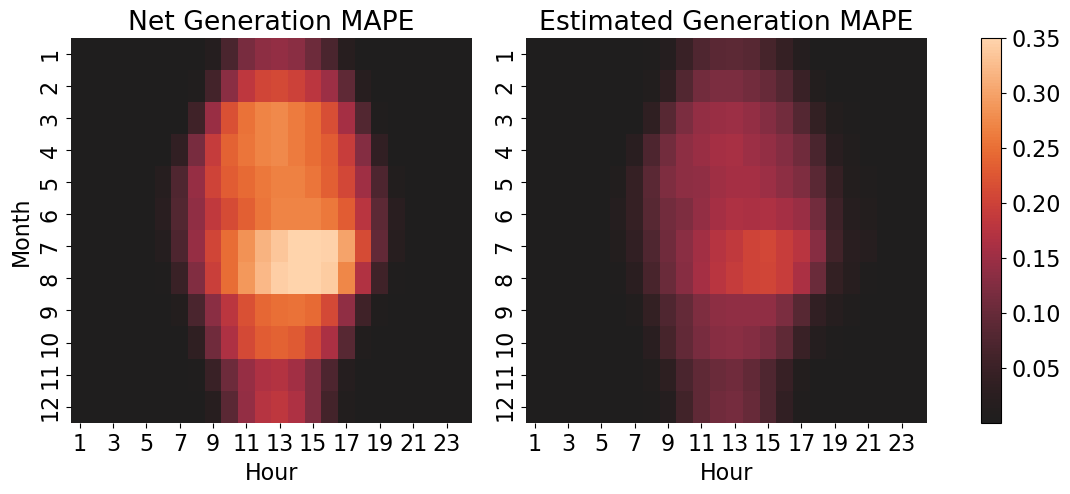}
    \caption{The absolute percent error, averaged across customers, is higher for the metered net generation compared with the estimated generation. The difference between the two extends to 15\% of installed capacity during peak hours. Positive error indicates the net generation MAPE is higher than the estimated generation MAPE.}
    \vspace{-0.2in}
    \label{fig:heatmap}
\end{figure}

The diurnal and seasonal impact of this approach is shown in Fig.\ \ref{fig:heatmap}. The MAPE of the net generation signal is quite high across the year, especially in the summer months and in the middle of the day. Application of this BTM generation reconstruction approach reduces the MAPE in the peak hours from $35\%$ to $15\%$. While the error in the estimated generation has not been reduced to zero, it has been reduced substantially and with reliance only on general weather conditions. 

The purpose of calculating the total generation and total consumption for each SPO customer is to build scenarios of BTM solar penetration from aggregations of different customers. The methodology was introduced with SPO customers and can be applied to any customer with a net-meter profile.

\begin{figure}[thpb]
\begin{center}
\includegraphics[width = 0.45\textwidth]{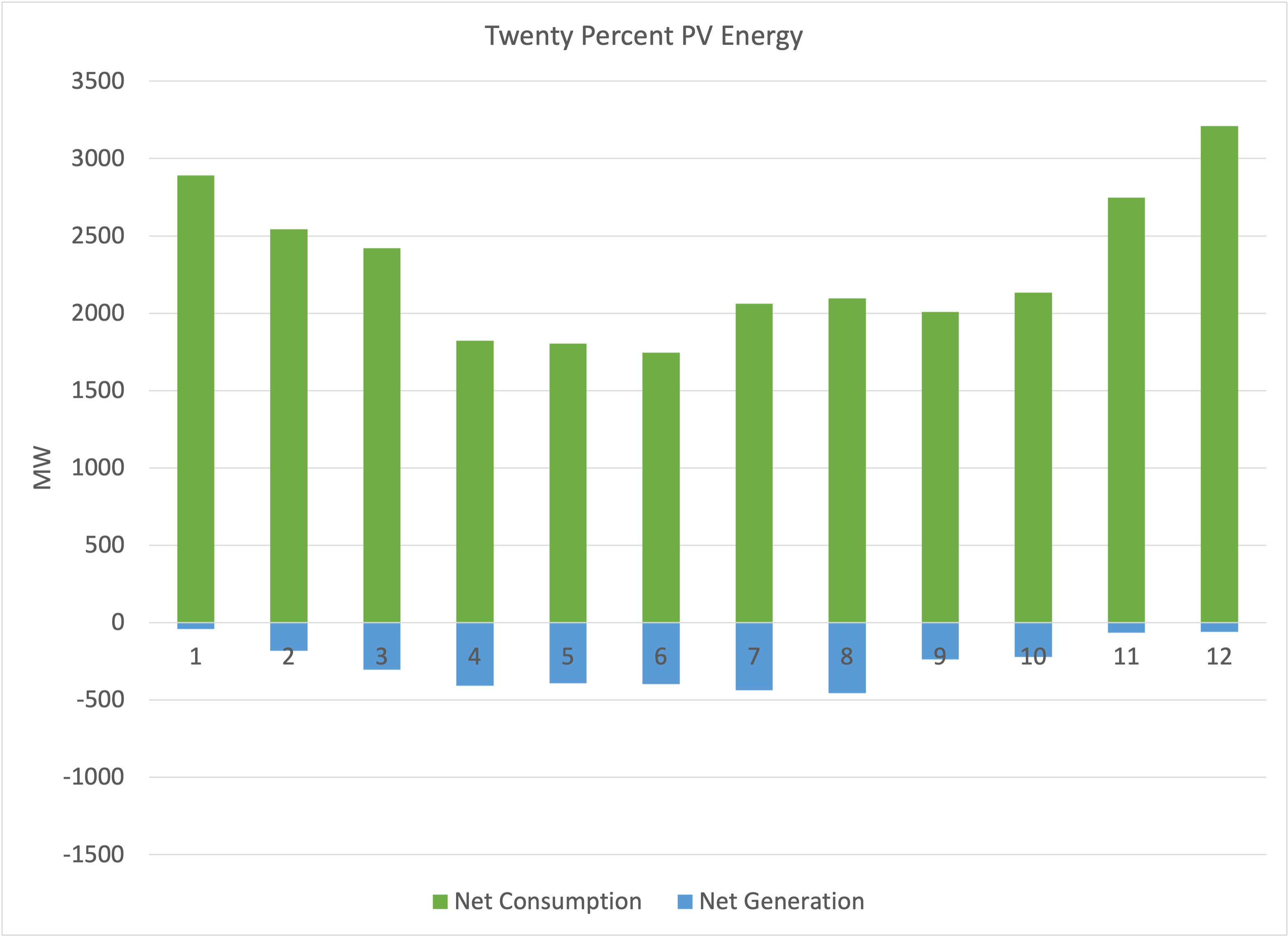}
\includegraphics[width = 0.45\textwidth]{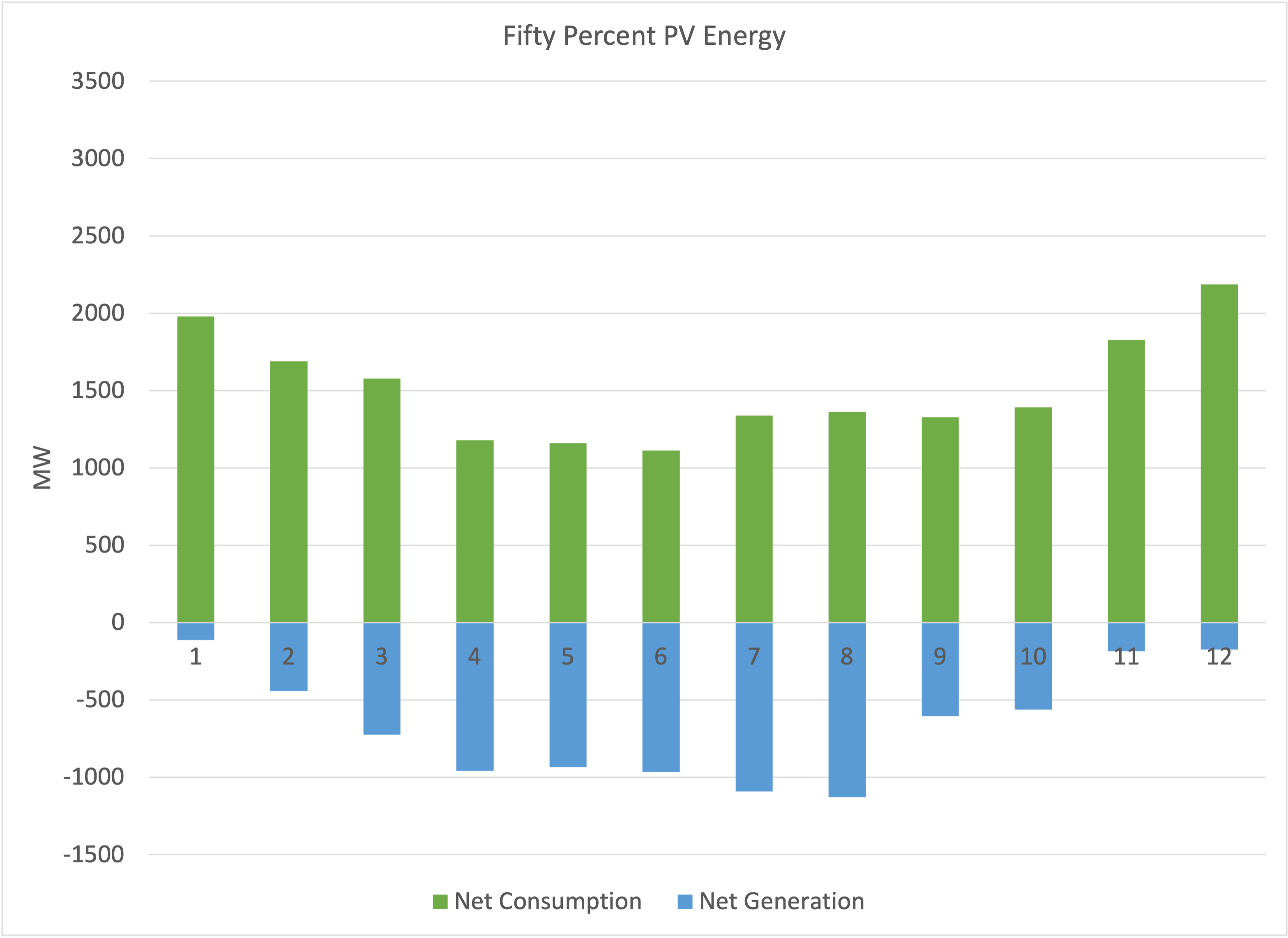}
\caption{Monthly consumption and generation for different BTM solar penetration levels: 20\% (top), 50\% (bottom).}
    \vspace{-0.2in}
\label{fig:twenty_thirty_fifty_monthly}
\end{center}
\end{figure}

The BTM  generation reconstruction algorithm has been applied to solar customers without a meter for total generation ($\sim36\%$ of customers in PGE, Table \ref{tab:PGE-cust}). Recalling that many solar customers size their systems to have no excess generation, the methodology described in this work will allow users to accurately represent BTM solar penetration and re-aggregate to artificial levels for testing of net-load forecasts. A demonstration of this re-aggregation to produce high penetrations of BTM solar service territories is shown in Fig.\ \ref{fig:twenty_thirty_fifty_monthly} for two scenarios. These figures are on the same vertical scale to emphasize the shift in net consumption (green, positive) to net generation (blue, negative).  

%
%

\section{Conclusion}\label{sec:concl}
 This work has introduced a methodology for creation of synthetic aggregations of customers to represent BTM solar penetration scenarios. The approach has focused on reconstruction of BTM solar generation; presence of other distributed energy resources (DERs) in the native load on rainy days would dampen the signal of true native load. Future work in improving the algorithm to tolerate ambiguity in DER beyond PV generation such as electric vehicles would require publicly available data that differentiates the DER signature from native load. The algorithm relies on publicly available weather condition observations to approximate interval native usage through distance matrix comparison with non-solar customers. This approach provides a more realistic measurement of the feeder-level BTM solar generation while remaining conservative in its estimation of interval generation. The algorithm can be used by net-load forecasters to create test scenarios of high penetration BTM solar against which they can test the robustness of their operational forecast to different adoption pathways. 

\section*{Acknowledgement}
The authors thank the SETO Technology Managers, Tassos Golnas and Kemal Celik, for their feedback and support.

\bibliographystyle{IEEEtran}
\bibliography{references}

\end{document}